\documentclass[aps,preprint,nofootinbib,floatfix]{revtex4}
\usepackage{amssymb}
\usepackage{graphicx}
\usepackage{amsmath}

\graphicspath{{./img/}}

\input{tcilatex}

\begin{document}

\title{Heterotic massive Einstein-Yang-Mills-type symmetry and Ward identity}
\author{Jen-Chi Lee}
\email{jcclee@cc.nctu.edu.tw}
\affiliation{Institute of Physics, National Chiao-Tung University, Hsin Chu, Taiwan, ROC}
\date{\today }

\begin{abstract}
We show that there exist spontaneously broken symmetries for massive modes
with transformation parameters $\theta _{\lbrack \mu \nu ]}^{(ab)},\theta
_{\mu }^{(ab)}$ etc. containing both Einstein and $E_{8}\otimes E_{8}$ (or $%
SO(32)$) Yang-Mills indices in the 10D Heterotic string. The corresponding
on-shell Ward identities are also constructed.
\end{abstract}

\maketitle

%


It was believed that the 10D fundamental string theory posseses an infinite
space-time symmetry structure at high energies \cite{1}. The recently
developed toy 2D string model reveals that this is indeed the case \cite{2}.
This is also suggested from, for example, the string field theory formalism 
\cite{3} although we still have no satisfactory nonperturbative second
quantized closed string formulation. In this paper, we would like to address
this issue from the $first$ quantized supersymmetric $massive$ $\sigma $%
-model point of view. We choose to work on the 10D Heterotic string \cite{4}
as it is still the one of most phenomenological interest. There have been
very few articles in the literature discussing the bosonic massive $\sigma $%
-model \cite{5}, not to mention the supersymmetric one. This is mainly
because it contains nonrenormalizable terms, and thus makes the perturbative
renormalization group $\beta $-function calculation uncontrollable. In this
short note, however, we will use the nonperturbative weal field
approximation (WFA) \cite{1} instead of the usual perturbative loop $(\alpha
^{\prime })$ expansion scheme. This will avoid the difficulty of the
nonrenormalizability of the massive $\sigma $-model (at least, to the first
order WFA). Moreover, the results we obtain from WFA are valid even at high
energies $(\alpha ^{\prime }\rightarrow \infty )$, thus WFA is the
appropriate approximation scheme to study the high energy behavior of
string. We will first discuss symmetry of the massless mode in the
Yang-Mills sector aiming to relate the $E_{8}\otimes E_{8}$ or $SO(32)$
gauge symmetry to the right-moving gauge states of the spectrum. This will
include both the S-matrix approach and the $modified$ $weak$ $field$ $\sigma 
$-model approach. We then generalize both approaches to the first (with
appropriate GSO-like projection ) massive case. After deriving the \ massive
Ward identities corresponding to the massive right-moving gauge states of
the spectrum, we will also construct explicit form of the weak field
symmetry transformation laws. $We$ $find$ $that$ $the$ $symmetry$ $%
transformation$ $parameters$ $contain$ $\theta _{\lbrack \mu \nu
]}^{(ab)},\theta _{\mu }^{(ab)}$ $etc.$ $with$ $mixed$ $Einstein$ $and$ $%
E_{8}\otimes E_{8}$ (or $SO(32)$) $Yang-Mills$ $indices,$ $which$ $point$ $%
field$ $theories$ $would$ $never$ $have$. This reveals the peculiar
structure of Heterotic string theory at high energies in contrast to the
renormalizable point field theories.

We fist discuss the massless Ward identity in the Yang-Mills sector of the
Heterotic string. The vertex operator for the massless gauge bosons in the
(0) and (-1) ghost pictures are%
\begin{equation}
V_{0}=\varepsilon _{\mu }^{a}V_{0}^{\prime \mu }(z)\overline{V}^{a}(%
\overline{z})=\varepsilon _{\mu }^{a}(\partial X^{\mu }+ik\cdot \psi \psi
^{\mu })\overline{J}^{a}(\overline{z})e^{ikX(z,\overline{z})},  \label{1aa}
\end{equation}%
\begin{equation}
V_{-1}=\varepsilon _{\mu }^{a}V_{-1}^{\mu }(z)\overline{V}^{a}(\overline{z}%
)=\varepsilon _{\mu }^{a}\psi ^{\mu }e^{-\phi }\overline{J}^{a}(\overline{z}%
)e^{ikX(z,\overline{z})},  \label{2aa}
\end{equation}%
with $k^{\mu }\varepsilon _{\mu }^{a}=0,k^{2}=0$. In Eqs. (\ref{1aa}) and (%
\ref{2aa}), $\overline{J}^{a}$ form the adjoint representation of either
SO(32) or $E_{8}\otimes E_{8}$ Kac-Moody algebra and $\phi $ is the
bosonized superconformal ghost. The following Ward identity can be easily
verified after calculating the four point correlator \cite{6}:%
\begin{equation}
k_{\mu }T^{\lambda a\mu b\nu c\alpha d}=k_{\mu }\int \overset{4}{\underset{%
i=1}{\prod }}d^{2}x_{i}<V_{0}^{\prime \lambda }(x_{1})\overline{V}%
^{a}(x_{1})V_{0}^{\prime \mu }(x_{2})\overline{V}^{b}(x_{2})V_{-1}^{\nu
}(x_{3})\overline{V}^{c}(x_{3})V_{-1}^{\alpha }(x_{4})\overline{V}%
^{d}(x_{4})>=0.  \label{3aa}
\end{equation}%
Note that we have chosen the ghost pictures for the vertex operators such
that the total superconformal ghost charge adds up to -2. The momentum $%
k_{\mu }$ is arbitrary chosen to be the momentum of the second vertex
operator. Eq.(\ref{3aa}) can be interpreted as the result of a gauge state
in the spectrum. In fact, there are two types of gauge states in the
right-moving NS-sector of the spectrum, they are \ (we use the notation in 
\cite{7})

Type I: 
\begin{equation}
G_{-1/2}|\chi >,\text{ where }G_{1/2}|\chi >=G_{3/2}|\chi >=L_{0}|\chi >=0,
\label{4aa}
\end{equation}

Type II:%
\begin{equation}
G_{-3/2}+2G_{-1/2}L_{-1}|\xi >,\text{ where }G_{1/2}|\xi >=G_{3/2}|\xi
>=(L_{0}+1)|\xi >=0.  \label{5aa}
\end{equation}%
Note that type I states have zero-norm at any space-time dimension while
type II states have zero-norm $only$ at D=10. The following massless gauge
state solution of Eq.(\ref{4aa})%
\begin{equation}
k_{\mu }b_{-1/2}^{\mu }\overline{J}^{a}|0,k>,  \label{6aa}
\end{equation}%
is the origin of the Ward identity (\ref{3aa}). We now discuss the symmetry
corresponding to the Ward identity in (\ref{3aa}) from modified
supersymmetric $\sigma $-model point of view \cite{1}. Instead of using the $%
\sigma $-model approach starting from a worldsheet action, we will write
down directly the weak field energy-momentum tensor $T$ and the supercurrent 
$T_{F}$ based on the consideration of vertex operators. Assume that $T$ and $%
T_{F}$ have the following weak field expansion:%
\begin{equation}
T=T^{(0)}+T^{(1)}+...;\text{ }T_{F}=T_{F}^{(0)}+T_{F}^{(1)}+...,  \label{7aa}
\end{equation}%
and the similar ones for the left-moving pieces. In Eq.(\ref{7aa}), each $%
T^{(0)}$ represents the corresponding free piece and each $T^{(1)}$ contains
the weak background fields. An $on$-$fixed$-$point$ $superconformal$ $%
deformation$ $(T^{(1)},\overline{T}^{(1)},$ $T_{F}^{(1)},\overline{T}%
_{F}^{(1)})$ is defined to be a deformation such that after deforming, the
superconformal charge algebra for $T^{\prime }$s is the same as that of $%
T^{(0)\prime }$s \cite{1}. It can be proved that a convenient choice to
satisfy the above conditions are $T^{(1)}=\overline{T}^{(1)},\overline{T}%
_{F}^{(0)}=\overline{T}_{F}^{(1)}=0$ (for the gauge sector) and $T_{F}^{(1)}$
has conformal dimension (1/2,1) w.r.t. $T^{(0)}$ and $\overline{T}^{(0)}$ 
\cite{1}. Now the superconformal deformation constructed from Eq.(\ref{1aa})
is \cite{1}%
\begin{equation*}
T^{(1)}=A_{\mu }^{a}\partial X^{\mu }\overline{J}^{a}+\partial _{\lbrack \nu
}A_{\mu ]}^{a}\psi ^{\nu }\psi ^{\mu }\overline{J}^{a},
\end{equation*}%
\begin{equation}
T_{F}^{(1)}=\frac{1}{2}A_{\mu }^{a}\psi ^{\mu }\overline{J}^{a},  \label{8aa}
\end{equation}%
with conditions%
\begin{equation}
\square A_{\mu }^{a}=0,\text{ }\partial ^{\mu }A_{\mu }^{a}=0.  \label{9aa}
\end{equation}%
Eq.(\ref{9aa}) is the conditions of vanishing renormalization group $\beta $%
-function of supersymmetric $\sigma $-model in the first order WFA and can
be interpreted as the linearized Yang-Mills equation in the covariant gauge.
On the other hand, the superconformal deformation constructed from gauge
state (\ref{6aa}) is%
\begin{equation}
\delta T=\partial _{\mu }\theta ^{a}\partial X^{\mu }\overline{J}^{a},\text{ 
}\delta T_{F}=\frac{1}{2}\partial _{\mu }\theta ^{a}\psi ^{\mu }\overline{J}%
^{a},  \label{10aa}
\end{equation}%
with condition $\square \theta ^{a}=0$. These induce the linearized form of
the $E_{8}\otimes E_{8}$ or $SO(32)$ gauge transformation in the covariant
gauge%
\begin{equation}
\delta A_{\mu }^{a}=\text{ }\partial _{\mu }\theta ^{a}.  \label{11aa}
\end{equation}%
We have thus related the massless Yang-Mills gauge symmetry of the Heterotic
string to the right-moving gauge state in the spectrum. One can get a
similar result for the gravitational sector. Note that the gauge conditions
in Eq.(\ref{9aa}) and gauge condition of $\theta ^{a}$ in Eq.(\ref{10aa})
are due to the on-shell formulation of the first quantized string. However,
it was shown how to dispense with these gauge conditions in Ref. \cite{8}.
Eq.(\ref{11aa}) is valid to all orders in $\alpha ^{\prime }$ in contrast to
the usual $\alpha ^{\prime }$ four-loop calculation \cite{9}. It does not
contain the usual homogeneous piece $C^{abc}A_{\mu }^{b}\theta ^{c}$.
However, this should not be thought of as a drawback of the formalism as has
been pointed out in \cite{1}. In fact, there are infinitely many terms in $%
string$ $theory$ in contrast to the usual Yang-Mills theory if one wants to
include them \cite{10}. Moreover, the advantage of the WFA formalism above
is that one can generalize the calculation to massive particles as we now
turn to discuss.

We first discuss the massive on-shell Ward identities. We will arbitrarily
choose two particles in the first massive Yang-Mills sector to illustrate
the identities \cite{7}. The vertex operators of them in the (0) ghost
picture read%
\begin{eqnarray}
V_{0}^{(1)} &=&[\varepsilon _{\lbrack \mu \nu \lambda ]}^{(ab)}(3\partial
X^{\mu }+ik\psi \psi ^{\mu })\psi ^{\nu }\psi ^{\lambda }-i\varepsilon
_{\lbrack \mu \nu ]}^{(ab)}(-\psi ^{\mu }\partial \psi ^{\nu }+ik\psi \psi
^{\mu }\partial X^{\nu })]\overline{J}^{(a}\overline{J}^{b)}(\overline{z}%
)e^{ikX(z,\overline{z})}  \notag \\
&\equiv &\varepsilon _{\lbrack \mu \nu \lambda ]}^{(ab)}V_{0}^{[\mu \nu
\lambda ]}\overline{V}^{(ab)}-i\varepsilon _{\lbrack \mu \nu
]}^{(ab)}V_{0}^{[\mu \nu ]}\overline{V}^{(ab)};  \notag \\
\varepsilon _{\lbrack \mu \nu ]}^{(ab)} &=&k^{\lambda }\varepsilon _{\lbrack
\lambda \mu \nu ]}^{(ab)},  \label{12aa}
\end{eqnarray}%
\begin{eqnarray}
V_{0}^{(2)} &=&[\varepsilon _{(\mu \nu )}^{(ab)}(\partial X^{\mu }\partial
X^{\nu }-\psi ^{\mu }\partial \psi ^{\nu }+ik\psi \psi ^{\mu }\partial
X^{\nu })-i/2\varepsilon _{\mu }^{(ab)}(\partial ^{2}X^{\mu }+ik\psi
\partial \psi ^{\mu })]\overline{J}^{(a}\overline{J}^{b)}e^{ikX(z,\overline{z%
})}  \notag \\
&\equiv &\varepsilon _{(\mu \nu )}^{(ab)}V_{0}^{(\mu \nu )}\overline{V}%
^{(ab)}-i/2\varepsilon _{\mu }^{(ab)}V_{0}^{\mu }\overline{V}^{(ab)};  \notag
\\
\varepsilon _{\mu }^{(ab)} &=&-k^{\nu }\varepsilon _{\nu \mu }^{(ab)},\text{ 
}\varepsilon _{\mu }^{(ab)}\text{ }^{\mu }-k^{\mu }k^{\nu }\varepsilon _{\nu
\mu }^{(ab)}=0.  \label{13aa}
\end{eqnarray}%
We have made the most general gauge choice for these two states. We are now
ready to calculate the correlator. For simplicity, we choose to calculate
three point correlator with two massless gauge vectors and one massive
state. The correlator of state (\ref{12aa}) decay into two vectors is
defined by the doublet $\{S^{\alpha a,[\mu \nu \lambda ]bc,\beta
d},S^{\alpha a,[\mu \nu ]bc,\beta d}\}$ where%
\begin{equation}
S^{\alpha a,[\mu \nu \lambda ]bc,\beta d}=\int \overset{3}{\underset{i=1}{%
\prod }}d^{2}x_{i}<V_{-1}^{\alpha }(x_{1})\overline{V}^{a}(\overline{x}%
_{1})V_{0}^{[\mu \nu \lambda ]}(x_{2})\overline{V}^{(bc)}(\overline{x}%
_{2})V_{-1}^{\beta }(x_{3})\overline{V}^{d}(\overline{x}_{3})>,  \label{14aa}
\end{equation}%
\begin{equation}
S^{\alpha a,[\mu \nu ]bc,\beta d}=-i\int \overset{3}{\underset{i=1}{\prod }}%
d^{2}x_{i}<V_{-1}^{\mu }(x_{1})\overline{V}^{a}(\overline{x}_{1})V_{0}^{[\mu
\nu ]}(x_{2})\overline{V}^{(bc)}(\overline{x}_{2})V_{-1}^{\beta }(x_{3})%
\overline{V}^{d}(\overline{x}_{3})>.  \label{15aa}
\end{equation}%
Similarly, the correlator of state (\ref{13aa}) decay into two vectors is
defined by the doublet $\{S^{\alpha a,(\mu \nu )bc,\beta d},S^{\alpha a,\mu
(bc),\beta d}\}$. The evaluation of Eqs.(\ref{14aa}) and (\ref{15aa}) is
straightforward. One can, for example, use the fermionic representation of
Heterotic string for the compatified left-moving sector%
\begin{equation}
\overline{J}^{a}=1/2:\overline{\lambda }^{i}\overline{\lambda }%
^{j}:(T^{a})^{ij};\text{ }Tr(T^{a}T^{b})=2\delta ^{ab},  \label{16aaa}
\end{equation}%
where $\overline{\lambda }^{i\prime }$s are 2D Majorana-Weyl Fermions and $%
T^{a}$ are the 496 generators of either $E_{8}\otimes E_{8}$ or SO(32). The
final results for Eqs. (\ref{14aa}) and (\ref{15aa}) are 
\begin{eqnarray}
S^{\alpha a,[\mu \nu \lambda ](bc),\beta d}
&=&24x_{12}x_{23}/x_{13}[Tr(T^{a}\{T^{b},T^{c}%
\}T^{d})+Tr(T^{a}T^{(b})Tr(T^{c)}T^{d})]  \notag \\
&&\times \{\eta ^{\alpha \lbrack \nu }\eta ^{\lambda |\beta }k_{1}^{|\mu
]}/x_{12}-\eta ^{\alpha \lbrack \nu }\eta ^{\lambda |\beta }k_{3}^{|\mu
]}/x_{23}\},  \label{17aa}
\end{eqnarray}%
\begin{eqnarray}
S^{\alpha a,[\mu \nu ](bc),\beta d} &=&16x_{12}x_{23}/x_{13}\{1/x_{12}(\eta
^{\alpha \lbrack \nu }\eta ^{\mu ]\beta }+k_{2}^{\alpha }k_{3}^{[\nu }\eta
^{\mu ]\beta }-k_{2}^{\beta }k_{3}^{[\nu }\eta ^{\mu ]\alpha })  \notag \\
&&+1/x_{23}(\eta ^{\alpha \lbrack \mu }\eta ^{\nu ]\beta }+k_{2}^{\beta
}k_{3}^{[\nu }\eta ^{\mu ]\alpha }-k_{2}^{\alpha }k_{3}^{[\nu }\eta ^{\mu
]\beta })\}  \notag \\
&&\times \lbrack
Tr(T^{a}\{T^{b},T^{c}\}T^{d})+Tr(T^{a}T^{(b})Tr(T^{c)}T^{d})]  \label{18aa}
\end{eqnarray}%
where $x_{ij}=\left\vert x_{i}-x_{j}\right\vert $. In deriving Eqs.(\ref%
{17aa}) and (\ref{18aa}), we have done the SL(2,C) gauge fixing and have
kept three positions of vertex operators at arbitrary but fixed values. Now
one of the first massive even G-parity gauge state solution of Eq.(\ref{4aa}%
) is%
\begin{equation*}
\{k_{\lambda }\theta _{\mu \nu }^{(ac)}b_{-1/2}^{\lambda }b_{-1/2}^{\mu
}b_{-1/2}^{\nu }+2\theta _{\mu \nu }^{(ac)}\alpha _{-1}^{\mu }b_{-1/2}^{\nu
}\}\overline{J}^{(a}\overline{J}^{c)}|0,k>,
\end{equation*}%
\begin{equation}
\theta _{\mu \nu }^{(ac)}=-\theta _{\nu \mu }^{(ac)},\text{ \ \ }k^{\mu
}\theta _{\mu \nu }^{(ac)}=0.  \label{19aa}
\end{equation}%
One can thus verify the following Ward identity after some algebra 
\begin{equation}
k_{2\lambda }\theta _{\mu \nu }^{(bd)}S^{\alpha a,[\mu \nu \lambda
](bd),\beta d}+2\theta _{\mu \nu }^{(bd)}S^{\alpha a,[\mu \nu ](bd),\beta
c}=0.  \label{20aa}
\end{equation}

We are now in a position to discuss the background field symmetry
transformation as we did for the massless mode. The on-fixed-point
superconformal deformations constructed from Eq. (\ref{12aa}) are%
\begin{eqnarray}
T^{(1)} &=&[M_{[\mu \nu \lambda ]}^{(ab)}(3\partial X^{\mu }+\overleftarrow{%
\partial }_{\alpha }\psi ^{\alpha }\psi ^{\mu })\psi ^{\nu }\psi ^{\lambda
}+E_{[\mu \nu ]}^{(ab)}(-\psi ^{\mu }\partial \psi ^{\nu }+\overleftarrow{%
\partial }_{\lambda }\psi ^{\lambda }\psi ^{\mu }\partial X^{\nu })]%
\overline{J}^{(a}\overline{J}^{b)},  \notag \\
T_{F}^{(1)} &=&\{3/2M_{[\mu \nu \lambda ]}^{(ab)}+2\partial ^{\alpha
}\partial _{\lambda }M_{[\mu \nu \alpha ]}^{(ab)}+1/2\partial _{\lambda
}E_{[\mu \nu ]}^{(ab)}\}\psi ^{\mu }\psi ^{\nu }\psi ^{\lambda }\overline{J}%
^{(a}\overline{J}^{b)}  \notag \\
&&+(3\partial ^{\lambda }M_{[\mu \nu \lambda ]}^{(ab)}+\partial ^{\lambda
}\partial _{\nu }E_{[\lambda \mu ]}^{(ab)}+E_{[\mu \nu ]}^{(ab)})\partial
X^{\mu }\psi ^{\nu }\overline{J}^{(a}\overline{J}^{b)};  \notag \\
E_{[\mu \nu ]}^{(ab)} &=&\partial ^{\lambda }M_{[\lambda \mu \nu ]}^{(ab)},%
\text{ \ }(\Box -2)M_{[\lambda \mu \nu ]}^{(ab)}=0.  \label{21aa}
\end{eqnarray}%
\bigskip $T_{F}^{(1)}$ in Eq. (\ref{21aa}) is a (1/2,1) superconformal
deformation as in Eq. (\ref{8aa}). The symbol $\leftarrow $ in Eq. (\ref%
{21aa}) means that $\overleftarrow{\partial }_{\alpha }$ acts backward on
the background fields. Note that $E_{[\mu \nu ]}^{(ab)}$ is the gauge
artifact of the propagating background field $M_{[\lambda \mu \nu ]}^{(ab)}$%
. If we compare Eq. (\ref{21aa}) with the superconformal deformation
constructed from Eq. (\ref{19aa}), we get the following symmetry
transformation:%
\begin{equation}
M_{[\mu \nu \lambda ]}^{(ab)}=\partial _{\lbrack \mu }\text{ }\theta _{\nu
\lambda ]}^{(ab)},  \label{22aa}
\end{equation}%
with conditions $(\Box -2)\theta _{\lbrack \mu \nu ]}^{(ab)}=\partial ^{\mu
}\theta _{\lbrack \mu \nu ]}^{(ab)}=0$. In deriving Eq.(\ref{22aa}) we have
used the gauge condition in Eq.(\ref{21aa}). Eq.(\ref{22aa}) is the symmetry
transformation corresponding to the Ward identity in Eq. (\ref{20aa}). It is
valid to all orders in $\alpha ^{\prime }$. Other symmetries and Ward
identities can be similarly constructed. For example, there exists a type II
symmetry transformation containing a nonderivative piece 
\begin{equation}
\delta G_{(\mu \nu )}^{(ab)}=2\partial _{\mu }\text{ }\partial _{\nu }\theta
^{(ab)}-\eta _{\mu \nu }\theta ^{(ab)},  \label{26aa}
\end{equation}%
where $(\Box -2)\theta ^{(ab)}=0$. We have used the appropriate gauge
conditions similar to Eq. (\ref{21aa}) to derive (\ref{26aa}). Those
symmetries which are generated by $massive$ gauge states, like Eqs.(\ref%
{22aa}) and (\ref{26aa}), were shown to be broken spontaneously by the
space-time metric $<\eta _{\mu \nu }>\neq 0$ \cite{10}.

It is possible to exhaust all solutions Eqs. (\ref{4aa}) and (\ref{5aa})
level by level. One then expects the existence of infinite number of mixed
high spin symmetry transformation parameters in the theory. Unfortunately,
the algebraic structure of this symmetry is still difficult to identify. The
situation is, however, much easier to handle in the case of toy 2D string
model where $w_{\infty }$ symmetry \cite{2} is believed to be related to the
discrete gauge states in the spectrum. Works in this direction is in
progress.

\bigskip

This work was supported by National Science Council of the R.O.C. under
contract \#NSC 82-0208-M-009-064.

\bigskip



\end{document}